\documentclass[runningheads]{svmult}

\usepackage{makeidx}   
\usepackage{graphicx}  
\usepackage{subeqnar}  
\usepackage{multicol}  
\usepackage{physprbb}  
\makeindex             

\begin{document}
\title*{Structure of Magnetocentrifugal Disk-Winds: From the
Launching Surface to Large Distances}
\toctitle{Structure of Magnetocentrifugal Disk-Winds:
\protect\newline From the Launching Surface 
to Large Distances}
%
%
\titlerunning{Magnetocentrifugal Disk-Winds}
%
\author{Ruben Krasnopolsky\inst{1}
\and Zhi-Yun Li\inst{2}
\and Roger D. Blandford\inst{3} }
\authorrunning{Krasnopolsky et al.}
%
%
\institute{Department of Astronomy and Astrophysics, University of Chicago,
Chicago, IL 60637, USA
\and Department of Astronomy, University of Virginia, Charlottesville,
VA 22903, USA
\and Theoretical Astrophysics, Caltech, Pasadena, CA 91125, USA} 
\maketitle              

\begin{abstract}
Protostellar jets and winds are probably driven magnetocentrifugally from 
the surface of accretion disks close to the central stellar objects. The
exact launching conditions on the disk, such as the distributions of 
magnetic flux and mass ejection rate, are poorly unknown. They could be 
constrained from observations at large distances, provided that a robust 
model is available to link the observable properties of the jets and winds
at the large distances to the conditions at the base of the flow. We 
discuss the difficulties in constructing such large-scale wind models,
and describe a novel technique which enables us to numerically follow 
the acceleration and propagation of the wind from the disk surface to 
arbitrarily large distances and the collimation of part of the wind 
into a dense, narrow ``jet'' around the rotation axis. Special attention
is paid to the shape of the jet and its mass flux relative to that of 
the whole wind. The mass flux ratio is a measure of the jet formation 
efficiency.
\end{abstract}

\section{Basic Mechanism and Previous Work}
The magnetocentrifugal mechanism is the leading candidate for producing 
the jets and winds observed around young stellar objects. The basic 
principle is relatively simple, and has been understood for
a long time~\cite{ref1}. It envisions parcels of fluid
element being lifted off and accelerated centrifugally along rapidly
rotating open magnetic field lines anchored firmly on an accretion disk.
Beyond a
certain point where the energy densities in the bulk flow motion and 
magnetic field are comparable, the field lines can no longer enforce 
rigid rotation, and the field becomes increasingly toroidal. It is the
``hoop stress'' associated with the toroidal field that is thought to 
be responsible for the wind collimation and jet production. The 
quantitative properties of the jet expected from this mechanism 
are poorly determined however, even though the
MHD equations that govern the wind structure and jet formation are
well known. Our understanding of the jet properties has been hampered 
to a large extent by the mathematical difficulties associated with 
obtaining wind solutions.

\subsection{Time-Independent Wind Solutions} 
There are two basic approaches in obtaining wind solutions. The first is to 
solve for the steady-state wind structure directly from the time-independent 
MHD equations. These equations can be cast into a second order partial 
differential equation (the Grad-Shafranov equation). It is well known 
that, for a cold wind that we are interested in here, the equation changes 
its type from being elliptic inside the fast (magnetosonic) surface to 
hyperbolic outside.
Computationally, the structure of the inner part of the wind inside the 
fast surface can be solved first by relaxation, and that beyond the fast 
surface later by the method of characteristics~\cite{ref2,ref3}.
The fact that the position of the fast surface, where the 
poloidal flow speed matches the fast magnetosonic speed, is unknown a 
priori poses a problem. To obtain a converged solution, one needs to have 
a good initial guess of the fast surface position, which is generally
difficult to obtain.

\subsection{Time-Dependent Numerical Simulations}
A more flexible approach is to numerically follow the time evolution of 
a wind to steady-state, if such a state exists. This approach has been
taken by several groups~\cite{ref4,ref5,ref6}. It is also the approach 
that we took~\cite{ref7}. Our simulations are based on the ZEUS MHD code,
and treat the Keplerian disk as a (lower) boundary, on which an open 
magnetic field of a prescribed distribution is anchored and from which 
cold material is injected into the wind at a prescribed rate. A novel 
feature of our simulation is the treatment of the region near the rotation 
axis, where the magnetocentrifugal mechanism is ineffective. The reason
is that to launch a cold parcel of fluid element centrifugally from a 
Keplerian disk the field line must incline an angle of at least $30^\circ$ 
away from the disk normal~\cite{ref1}. This condition is not 
met in the axial region since the field line along the axis must be 
exactly vertical by symmetry. In reality, the axial region may be filled
with a normal stellar wind from the central object or bundles of open 
field lines from the stellar magnetosphere. We are thus motivated to 
inject a light fluid with little mass flux at a speed fast enough to 
escape from the potential well along those field lines that fail to 
operate magnetocentrifugally. The light axial flow provides a plausible 
inner boundary to the magnetocentrifugal disk-wind, the focus of our 
investigation.

\begin{figure}
\begin{center}
\includegraphics[width=.4\textwidth]{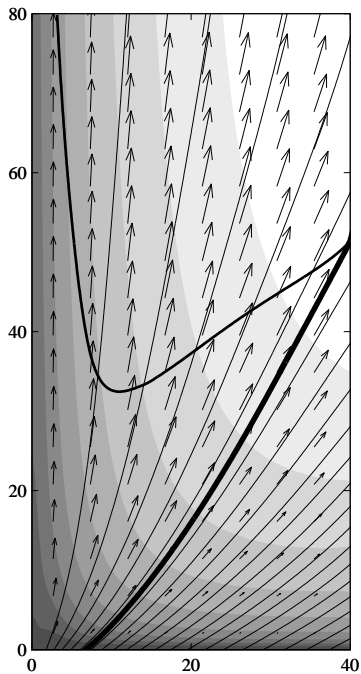}
\end{center}
\caption[]{A representative magnetocentrifugal wind launched from a
Keplerian disk (in arbitrary units; taken from~\cite{ref7}). Shown 
are the magnetic
field lines ({\it light solid lines}), velocity vectors ({\it arrows}),
density contours ({\it shades}), and the fast magnetosonic surface ({\it 
solid line of medium thickness}). The thickest solid line
divides the portion of the wind that becomes super fast-magnetosonic
inside the simulation box ({\it above}) from the portion that does not
({\it below}).  }
\label{fig1}
\end{figure}

A typical example of the steady-state disk-wind solutions obtained from 
time-dependent simulations is given in Fig.~\ref{fig1}. The wind is 
driven off all of the (equatorial) disk surface. Flow acceleration is
apparent along all field lines except those near the axis where a fast 
initial injection is imposed. Note that the field (and stream) lines 
collimate gradually, as expected. What was not expected was the great 
care that went into designing the shape of the simulation box, so that 
a steady state could be reached at all. If we were to cut the box shown 
in Fig.~\ref{fig1} in half or to elongate the box horizontally instead 
of vertically, and restart the simulation with the same initial and 
boundary conditions, the wind would become chaotic. The sensitive solution 
dependence on the simulation box has also been noted by others~\cite{ref5}.
It is a major concern for the time-dependent approach to finding disk-wind 
solutions.

\section{Magnetocentrifugal Winds From Inner Accretion Disks}
We believe that the simulation box dependence described above comes from 
the fact that a large fraction of the wind remains completely sub 
fast-magnetosonic in the computational domain, as shown in lower-right 
corner of Fig.~\ref{fig1}. Information on the sub-fast outer boundary 
can propagate upstream all the way to the disk surface and interfere 
with the wind launching. The reason for the region to remain sub 
fast is simple: the wind coming off the outer part of the disk encounters 
the edge of the simulation box too soon; it simply does not have enough 
room to get accelerated to the fast speed. This situation remains as 
long as the wind is driven off from {\it all} of the (equatorial) disk plane 
(as assumed in Fig.~\ref{fig1} and other previous time-dependent disk-wind 
simulations) regardless of the box size. It motivates us to restrict the 
wind launching to only the inner region of an accretion disk, and focus 
on inner-disk driven winds for which the simulation box dependence 
disappears.

\subsection{Inner-Disk Driven Winds: Simulation Setup}

\begin{figure}
\begin{center}
\includegraphics[width=.45\textwidth]{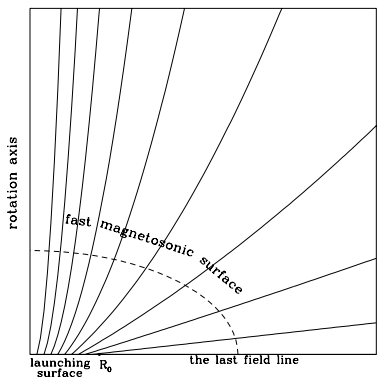}
\end{center}
\caption[]{Schematic view of a cold magnetocentrifugal wind launched 
{}from a limited, inner disk region}
\label{fig2}
\end{figure}

Physically, the wind launching may be limited to the inner region of 
a protostellar disk where the temperature is high enough (greater than 
$\sim 10^3$ K) that thermal ionization of alkali metals can provide
 enough charges to couple the magnetic field to the disk matter. For 
typical parameters, this occurs inside a radius of order 1 AU.
Numerically, we set up the simulation as sketched in Fig.~\ref{fig2}.
To fill all available space above (and below) the equatorial plane,
we demand the last field line anchored at the outer radius of
the launching region $R_0$ to lie exactly on the equatorial plane.
Wind plasma sliding along this last (horizontal) field line will become 
super fast-magnetosonic in the computational realm, provided that the 
size of the simulation box is sufficiently large. Once the whole fast 
surface is completely enclosed inside the simulation box, the size
and shape of the box would have minimal effects on the structure of 
the wind, especially near the launching surface, since information 
cannot propagate upstream in a super fast region. In this way, we should 
be able to study the wind structure up to arbitrarily large distances
{}from the source region, limited only by computer time.

\subsection{Large-Scale Wind Structure: Numerical Results}

\begin{figure}
\begin{center}
\includegraphics[width=.50\textwidth]{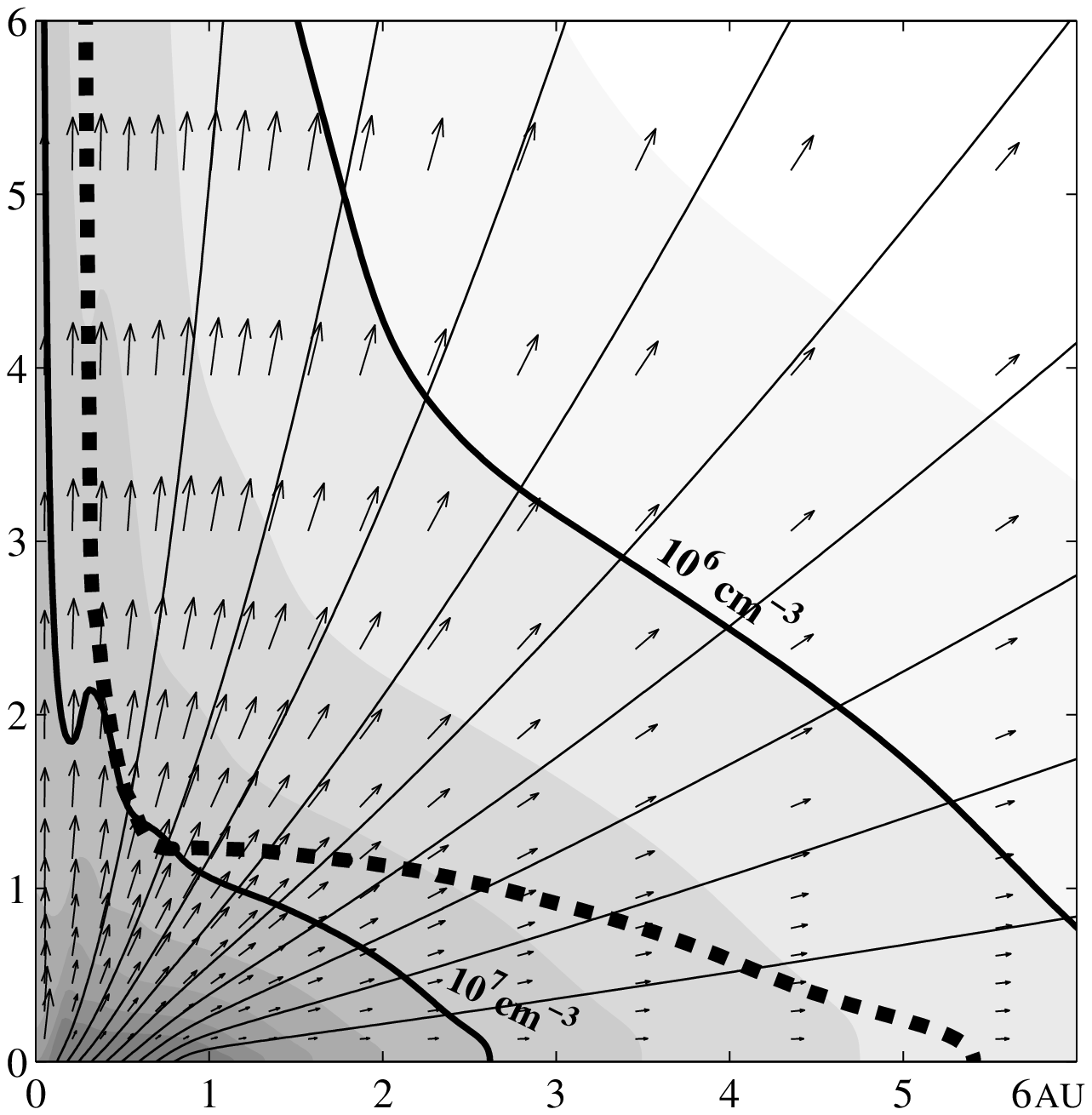}
\end{center}
\begin{center}
\includegraphics[width=.52\textwidth]{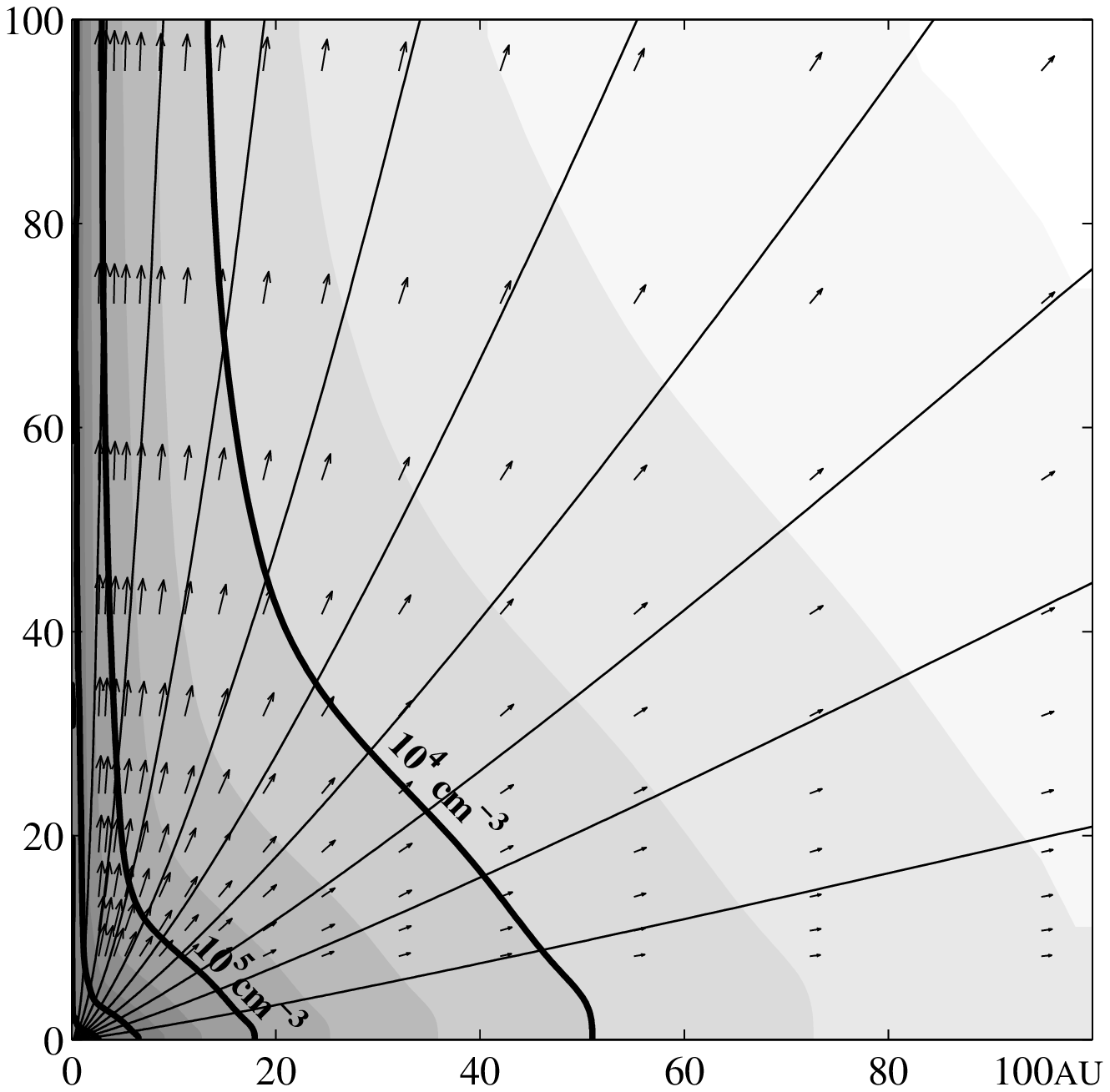}
\end{center}
\caption[]{Streamlines ({\it light solid lines}) and density contours 
({\it heavy solid lines and shades}) of a representative steady wind 
driven from the inner region of an protostellar disk in a 6 AU ({\it 
top panel}) and $10^2$ AU ({\it bottom panel}) box.
The fast surface ({\it dashed line}) is also plotted
in the smaller box. The streamlines
divide the wind into 10 zones of equal mass flux, with a total
mass flux of $10^{-8} M_\odot$yr$^{-1}$ per side of the disk.
The gray scale shows the log of the hydrogen number density,
with three shades per decade. The density contours correspond 
to $10^4$, $10^5$, $10^6$ and $10^7$ in units of cm$^{-3}$.}
\label{fig3}
\end{figure}

For illustration, we consider a specific example. We adopt as the launching 
conditions on the disk a power-law distribution of the vertical field 
strength with radius as $B_z\propto R^{-1.5}$ and a mass injection rate
(per unit area) $j\propto R^{-0.5}$ between 0.1 and 0.8 AU. The inner
radius is chosen to mimic the disk truncation radius due to the stellar
magnetosphere. Inside this radius, we inject a fast light flow as described 
earlier. At the edge of the wind launching region, taken to be $R_0=1$ AU
for simplicity, we impose the condition that $B_z=j=0$ since the last 
field (and stream) line must be horizontal. A cubic polynomial is used to 
connect smoothly the values of $B_z$ (or $j$) between 0.8 and 1 AU.
As before,
we follow the time evolution of the wind numerically to a steady state.
The steady wind solution, from the launching surface (inside 1 AU) all
the way to a large distance of $10^2$ AU, is displayed in Fig.~\ref{fig3}
on two scales.

Several features are worth noting. First, the fast surface shown in the 
smaller box closes on the equatorial plane, as advertised. This closure 
enabled us to continue the wind solution on to large distances without 
having to worry about the effects of box size. Second, the wind speed 
is anisotropic, with a value roughly 3 times higher in the axial region 
than in the equatorial region. The anisotropy appears to be even stronger
in density, which is stratified more or less cylindrically (or jet-like)
near the rotation axis, as expected. In the more equatorial region, the 
density contours bulge outward prominently, retaining some memory of 
the nearly horizontal shape of the initial density contour. This 
non-cylindrical shape of density contours is significant because the 
wind emission in forbidden lines such as [SII]$\lambda\lambda$6716,6731
is sensitive to the density~\cite{ref8}, and the shape of the jet may
resemble to a zeroth order the shape of the density contour at some
fiducial value. We choose to represent the outer boundary of a ``jet'' 
by a fiducial density contour of $10^4$ cm$^{-3}$ (the outermost contour 
in the larger box of Fig.~\ref{fig3}). The ``jet'' so defined has a width 
of $\sim 30$ AU at a height of $10^2$ AU, comparable to that observed in 
HH 30 jet. The bulging out at the ``jet'' base is not observed, however.
Furthermore, the ``jet'' contains only about a quarter of the total wind 
mass flux, making its formation rather inefficient. These undesirable
``jet'' features demonstrate that not all combinations of the launching 
conditions are capable of producing cylindrical jets that contain the
majority of the wind mass flux. We find that one way to improve the jet 
shape {\it and} increase its mass flux fraction is to make the mass 
injection rate $j$ on the disk decrease more steeper with radius. The 
details will be presented in a forthcoming paper.

\section{Conclusion}

To summarize, by limiting the wind launching to the inner part of an 
accretion disk, we are able to obtain using time-dependent 
simulation steady-state wind solutions that extend
{}from the launching surface 
to large distances. Combined with a detailed calculation of the
thermal structure and emission properties, these large-scale wind
solutions can be used to yield constraints on the launching 
conditions from the properties of jets and winds observed at the 
large distances. The constraints may provide clues to the origin
of the disk magnetic fields that launch the jets and winds.

%

\end{document}